\documentclass[conference]{IEEEtran}
\makeatletter
\def\ps@headings{%
\def\@oddhead{\mbox{}\scriptsize\rightmark \hfil \thepage}%
\def\@evenhead{\scriptsize\thepage \hfil \leftmark\mbox{}}%
\def\@oddfoot{}%
\def\@evenfoot{}}
\makeatother
\usepackage{graphicx,epstopdf,colortbl}
\usepackage{amsmath, amssymb,cite,comment}
\usepackage{enumerate}
\usepackage{times}
\usepackage{multirow,multicol}
\usepackage[ruled, vlined]{algorithm2e}

\newcounter{ctr}

\makeatletter

\newcommand{\be}{\begin{eqnarray}}
\newcommand{\ee}{\end{eqnarray}}
\newcommand{\nn}{\nonumber}

\newcommand{\bm}{\boldmath}

\newcommand{\mc}{\multicolumn}

\newcommand{\C}{\mbox{\bm $C$}}

\newcommand{\cc}{\mbox{\bm $c$}}
\newcommand{\m}{\mbox{\bm $m$}}

\newcommand{\uu}{\mbox{\bm $u$}}

\newcommand{\sss}{\mbox{\bm $s$}}

\newcommand{\gggg}{\mbox{\bm $g$}}

\newcommand{\0}{{\bf 0}}

\newcommand{\vv}{\mbox{\bm $v$}}

\newcommand{\yy}{\mbox{\bm $y$}}

\begin{document}

\title{Exact Regenerating Codes for Byzantine Fault
Tolerance in Distributed Storage}

\author{
 \begin{tabular}{ccc}
 Yunghsiang~S.~Han &  Rong Zheng & Wai Ho Mow\\
Dept. of Electrical Engineering & Dept. of Computer Science & Dept. of Electrical and Electronic Engineering\\
National Taiwan Univ. of Sci. and Tech. & University of Houston &  Hong Kong Univ. of Sci. and Tech.\\
Taiwan, R.O.C.  & Houston, TX &  Hong Kong \\
E-mail: {\it yshan@mail.ntust.edu.tw} & E-mail: {\it rzheng@uh.edu} &E-mail: {\it eewhmow@ust.hk}
\end{tabular}
}


\maketitle

\thispagestyle{plain}

\begin{abstract}
Due to the use of commodity software and hardware, crash-stop and Byzantine
failures are likely to be more prevalent in today's large-scale distributed
storage systems. Regenerating codes have been shown to be a more efficient way
to disperse information across multiple nodes and recover crash-stop failures
in the literature. In this paper, we present the design of regeneration codes in
conjunction with integrity check that allows exact regeneration of failed nodes
and data reconstruction in presence of Byzantine failures. A progressive
decoding mechanism is incorporated in both procedures to leverage computation
performed thus far. The fault-tolerance and security properties of the schemes
are also analyzed. 
\end{abstract}


\setcounter{page}{1}

\section{Introduction}
\label{sect:intro}
Storage is becoming a commodity due to the emergence of new storage media and
the  ever decreasing cost of conventional storage devices. Reliability, on the
other hand, continues to pose challenges in the design of large-scale
distributed systems such as data centers. Today's data centers operate on
commodity hardware and software, where both crash-stop and Byzantine failures
(as a result of software bugs, attacks) are likely the norm. To achieve
persistent storage, one common approach is to disperse information pertaining
to a data file (the message) across nodes in a network. For instance, with
$(n,k)$ maximum-distance-separable (MDS) codes such as Reed-Solomon (RS) codes,
data is encoded and  stored across $n$ nodes and, an end user or a data
collector can retrieve the original data file by accessing {\it any} $k$ of the
storage nodes, a process referred to as {\it data reconstruction}. 

Upon failure of any storage node,  data stored in the failed node needs to be
regenerated (recovered) to maintain the functionality of the system. A
straightforward way for data recovery is to first reconstruct the original data
and then regenerate the data stored in the failed node.  However, it is
wasteful to retrieve the entire $B$ symbols of the original file, just to
recover a small fraction of that stored in the failed node. A more efficient
way is to use the {\it regenerating codes} which was introduced in the pioneer
works by Dimakis {\it et al.} in ~\cite{DIM07,DIM10}.  A
tradeoff can be made between the storage overhead and the repair bandwidth
needed for regeneration.  Minimum Storage Regenerating (MSR) codes minimize
first, the amount of data stored per node, and then the repair bandwidth, while
Minimum Bandwidth Regenerating (MBR) codes carry out the minimization in the
reverse order. The design of regenerating codes have received much attention in
recent years~\cite{WU07,WU10,CUL09,WU09,RAS09,PAW11,OGG11,RAS11}. Most notably, Rashi
{\it et al.} proposed optimal exact-Regenerating codes using a product-matrix
reconstruction that recover exactly the same stored data of the failed node
(and thus the name exact-regenerating)~\cite{RAS11}.  Existing work assumes
crash-stop behaviors of storage nodes. However, with Byzantine failures, the
stored data may be tampered resulting in erroneous data reconstruction and
regeneration. 

In this paper, we consider the problem of exact regeneration for Byzantine
fault tolerance in distributed storage networks. Two challenging issues arise
when nodes may fail arbitrarily. First, we need to verify whether the
regenerated or reconstructed data is correct. Second, efficient algorithms are
needed that {\it incrementally} retrieve additional stored data and perform
data-reconstruction and regeneration when errors have been detected. Our work
is inspired by~\cite{RAS11} and makes the following new contributions:
\begin{itemize}
\item We present the detailed design of an exact-regenerating code with error
correction capability.\footnote{The encoding process is the same as that
given in~\cite{RAS11} except that an explicit encoding matrix is given in this
work.}
\item We devise a procedure that verifies the correctness of
regenerated/reconstructed data.
\item We propose progressive decoding algorithms for data-reconstruction and
regeneration that leverages computation performed thus far. 
\end{itemize}

The rest of the paper is organized as follows. We give an overview of
regenerating codes and RS codes in Section~\ref{sect:background} to prepare the
readers with necessary background. The design of error-correcting
exact regenerating code for the MSR points and MBR points are presented in
Section~\ref{sect:msr} and Section~\ref{sect:mbr}, respectively. Analytical results on the
fault tolerance and security properties of the proposed schemes are given in
Section~\ref{sect:eval}. Related work is briefly surveyed in
Section~\ref{sect:related}. Finally, we conclude the paper  in Section~\ref{sect:conclusion}. 

\section{Preliminaries}
\label{sect:background}
\subsection{Regenerating Codes}
Regenerating codes achieve bandwidth efficiency in the regeneration process by
storing additional symbols in each storage node or accessing more storage
nodes.  Let $\alpha$ be the number of symbols over finite field $GF(q)$ stored
in each storage node and $\beta\le\alpha$ the number of symbols downloaded from
each storage during regeneration.  To repair the stored data in the failed
node, a helper node accesses $d$ surviving nodes with the total repair
bandwidth $d\beta$. In general, the total  repair bandwidth is much less than
$B$. A regenerating code can be used not only to regenerate coded data but also
to reconstruct the original data symbols.  Let the number of storage nodes be
$n$.  An $[n,k,d]$ regenerating code requires at least $k$ and $d$ surviving
nodes to ensure successful data-reconstruction and
regeneration~\cite{RAS11}, respectively. Clearly, $k\le d\le n-1$.

The main results given in~\cite{WU07,DIM10} are the so-called cut-set bound on
the  repair bandwidth. It states that any regenerating code must satisfy
the following inequality: 
\begin{eqnarray}
B\le \sum_{i=0}^{k-1} \min\{\alpha,(d-i)\beta\}~.\label{main-inequality}
\end{eqnarray}
Minimizing $\alpha$ in~\eqref{main-inequality} results in a regenerating code
with minimum storage requirement; and minimizing $\beta$ results in that with
minimum repair bandwidth. It is impossible to have minimum values both on
$\alpha$ and $\beta$ concurrently, and thus there exists a tradeoff between
storage and repair bandwidth.  The two extreme points
in~\eqref{main-inequality} are referred as the minimum storage
regeneration (MSR) and minimum bandwidth regeneration (MBR) points,
respectively. The values of $\alpha$ and $\beta$ for MSR point can be obtained
by first minimizing $\alpha$ and then minimizing $\beta$:
\begin{eqnarray}
\alpha&=&\frac{B}{k}\nonumber\\
\beta&=&\frac{B}{k(k-d+1)}~.\label{MSR}
\end{eqnarray}
Reversing the order of minimization we have  $\beta$ and $\alpha$ for MBR as
\begin{eqnarray}
\beta&=&\frac{2B}{k(2d-k+1)}\nonumber\\
\alpha&=&\frac{2dB}{k(2d-k+1)}~.\label{MBR}
\end{eqnarray}
As defined in~\cite{RAS11},   an $[n,k,d]$ regenerating code with parameters
$(\alpha,\beta,B)$ is optimal if i) it satisfies the cut-set bound with equality,
and ii) neither $\alpha$ and $\beta$ can be reduced unilaterally without
violating the cut-set bound.  Clearly,  both MSR and MBR codes are optimal
regenerating codes.

It has been proved that when designing  $[n,k,d]$ MSR  or MBR codes, it
suffices to consider those with $\beta=1$~\cite{RAS11}. Throughout this paper,
we assume that $\beta=1$ for code design. Hence~\eqref{MSR} and \eqref{MBR}
become
\begin{eqnarray}
\alpha&=&d-k+1\nn\\
B&=&k(d-k+1)=k\alpha\label{NMSR}
\end{eqnarray}
and
\begin{eqnarray}
\alpha&=&d\nn\\
B&=&kd-k(k-1)/2~,\label{NMBR}
\end{eqnarray}
respectively, when $\beta=1$.

There are two ways to regenerate data for a failed node. If the replacement
data generated is exactly the same as those stored in the failed node, we call
it the {\it exact regeneration}. If the replacement data generated is only to
guarantee the data-reconstruction and regeneration properties, it is called
{\it functional regeneration}. In practice, exact regeneration is more desired
since there is no need to inform each node in the network regarding the
replacement. Through this paper, we only consider exact regeneration and design
exact-regenerating codes with error-correction capabilities. 

\subsection{Reed-Solomon codes}
\label{sec:rs}
\begin{figure*}[htp]
\begin{center}
\includegraphics[width=6in]{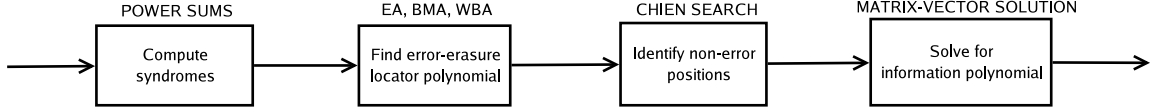}
\caption{Block diagram of RS decoding. Above each block, the corresponding existing algorithms are indicated.}
\label{fig:rs_decode}
\end{center}
\end{figure*}

Since Reed-Solomon (RS) codes will be used in the design of regenerating
codes, we briefly describe the encoding and decoding mechanisms of RS codes
next.

RS codes are the most well-known error-correction codes. They not only can
recover data when nodes fail, but also can guarantee recovery when a subset of
nodes are Byzantine. RS codes operate on symbols of $m$ bits, where all symbols
are from finite field $GF(2^m)$.  An $[n,d]$ RS code is a linear code, with
parameters $n=2^{m}-1$ and $n-d=2t~,$ where $n$ is the total number of symbols
in a codeword, $d$ is the total number of information symbols, and $t$ is the
symbol-error-correction capability of the code.
\paragraph*{Encoding}
Let the sequence of $d$ information symbols in $GF(2^{m})$ be
$\uu=[u_0,u_1,\ldots,u_{d-1}]$ and $u(x)$ be the information polynomial of $\uu$
represented as\footnote{We use polynomial and vectorized representations of information symbols, codewords, received symbols and errors interchangeably in this work.}
$$u(x)=u_0+u_1x+\cdots+u_{d-1}x^{d-1}~.$$
The codeword polynomial, $c(x)$, corresponding to $u(x)$ can be
encoded as
\begin{eqnarray}
c(x)=u(x)x^{n-d}+(u(x)x^{n-d}\mod g(x))~,\label{encoder-1}
\end{eqnarray}
where $g(x)$ is a generator polynomial of the RS code. It is
well-known that $g(x)$ can be obtained as
\begin{eqnarray}
\label{g(x)}
g(x)&=&(x-a^b)(x-a^{b+1})\cdots(x-a^{b+2t-1})\nonumber\\
&=&g_0+g_1x+g_2x^2+\cdots+g_{2t}x^{2t}~,
\end{eqnarray}
where $a$ is a generator (or a primitive element) in $GF(2^{m})$, $b$ an arbitrary integer, and $g_i\in
GF(2^{m})$. The RS code defined by~(\ref{encoder-1}) is a systematic code, where the information symbols $u_0,u_1,\ldots,u_{d-1}$ occur as coefficients (symbols) in $c(x)$.

Another encoding method for RS codes is the encoder proposed by  Reed and Solomon~\cite{REE60}, where the codeword $\cc$ corresponding to the information sequence $\uu$ is
\begin{eqnarray}
\cc=[u(a^0),u(a^1),u(a^2),\cdots,u(a^{n-1})]~.\label{encoder-2}
\end{eqnarray}
When $b=1$,  the codes generated by (\ref{encoder-1}) and (\ref{encoder-2}) are
identical. In this work, we adopt the later encoding method.

\paragraph*{Decoding}
The decoding process of RS codes is more complex. A complete description can be
found in~\cite{MOO05}.

Let $r(x)$ be the received polynomial and
$r(x)=c(x)+e(x)+\gamma(x)=c(x)+\lambda(x)$, where $e(x)=
\sum_{j=0}^{n-1}e_jx^j$ is the error polynomial, $\gamma(x)=
\sum_{j=0}^{n-1}\gamma_jx^j$  the erasure polynomial, and
$\lambda(x)=\sum_{j=0}^{n-1}\lambda_jx^j=e(x)+\gamma(x)$  the
errata polynomial. Note that $g(x)$ and (hence) $c(x)$ have
$\alpha^b,\alpha^{b+1},\ldots,\alpha^{b+2t-1}$ as roots. This property is used
to determine the error locations and recover the information symbols.

The RS codes are optimal as it provides the largest separation among code
words, and an $[n, d]$ RS code can recover from any $v$ errors as long as $v
\le \lfloor{\frac{n-d-s}{2}}\rfloor$, where $s$ is the number of erasure (or
irretrievable symbols).  The basic procedure of RS decoding is shown in
Figure~\ref{fig:rs_decode}. The last step in this figure is not necessary if a
systematic RS code is applied; otherwise,  the last step of the decoding
procedure involves solving a set of linear equations, and can be made efficient
by the use of Vandermonde generator matrices~\cite{william1988numerical}. The decoding that handles both error and erasure is called the error-erasure decoding.

In $GF(2^m)$, addition is equivalent to bit-wise exclusive-or (XOR), and
multiplication is typically implemented with multiplication tables or discrete
logarithm tables.  To reduce the complexity of multiplication, Cauchy
Reed-Solomon (CRS) codes~\cite{Blomer95anxor-based} have been proposed to use a
different construction of the generator matrix, and convert multiplications to
XOR operations for erasure.  However, CRS codes incur the same complexity as RS
codes for error correction. 

\section{Encoding and Decoding of Error-Correcting Exact-Regenerating Codes for the MSR Points}
\label{sect:msr}
In this section, we demonstrate how to perform error correction on MSR codes
designed to handle Byzantine failures by extending the code construction
in~\cite{RAS11}. It has been proved in~\cite{RAS11} that an MSR code $\C'$ with
parameters $[n',k',d']$ for any $2k'-2\le d'\le n'-1$ can be constructed from
an MSR code $\C$ with parameters $[n=n'+i,k=k'+i, d=d'+i]$, where $d=2k-2$ and
$i=d'-2k'+2$. Furthermore, if $\C$ is linear, so is $\C'$. Hence, it is
sufficient to design an MSR code for $d=2k-2$. When $d=2k-2$ we have
$$\alpha=d-k+1=k-1=d/2$$ and $$B=k\alpha=\alpha(\alpha+1)~.$$ We assume that
the  symbols in  data are elements from $GF(2^m)$. Hence, the total data in
bits is $mB$ bits for $\beta=1$.
\subsection{Verification for Data-Reconstruction}
\label{sec:verification-data}
Since we need to design codes with Byzantine fault tolerance it is necessary to
perform integrity check after the original data is reconstructed.
Two common verification mechanisms can be used: CRC  and hash function.
Both methods add redundancy to the original data before they are encoded.  Here
we adopt CRC since it is simple to implement and requires less
redundancy. 

CRC  uses a cyclic code (CRC code) such that each information
sequence can be verified using its generator polynomial with degree $r$,
where $r$ is the redundant bits added to the information
sequence~\cite{MOO05,REE99}.  The amount of errors that can be detected by
a CRC code is related to the number of redundant bits. A CRC code with $r$
redundant bits {\it cannot} detect $(\frac{1}{2^r})100\%$ portion of errors or
more.  For example, when $r=32$, the mis-detection error probability is on the
order of $10^{-10}$. Since the size of original data  is usually large, the
redundancy added by imposing a CRC code is  relatively small. For example,
for a $[100,20,38]$ MSR code with $\alpha=19,\ B=19\times 20=380$, we need to
operate on $GF(2^{11})$ such that the total bits for original data are $4180$.
If $r=32$, then only $0.77\%$ redundancy is added. Hence, in the following, we
assume that the CRC checksum has been added to the original data  and the resultant size
is $B$ symbols.

\subsection{Encoding}
\label{sec:encoding}
We arrange the information sequence $\m=[m_0,m_1,\ldots, m_{B-1}]$ into
an information vector $U$ with size $\alpha\times d$ such that
\begin{eqnarray*}
u_{ij}=\left\{\begin{array}{cc}
u_{ji}=m_{k_1}&\mbox{ for } i\le j\le \alpha\\
u_{(j-\alpha)i}=m_{k_2}&\mbox{ for } i+\alpha\le j\le 2\alpha
\end{array}\right.~,
\end{eqnarray*}
where $k_1=(i-1)(\alpha+1)-i(i+1)/2+j$ and $k_2=(\alpha+1)(i-1+\alpha/2)-i(i+1)/2+(j-\alpha)$.
Let $U=\left[A_1A_2\right]$. From the above construction, $A_j$'s are symmetric matrix with 
dimension $\alpha\times\alpha$ for $j=1,2$.  

In this encoding, each row of for the information vector $U$ produces a
codeword of length $n$. An $[n,d=2\alpha]$ RS code is adopted to construct the
MSR code. In particular, for the $i$th row of $U$, the corresponding codeword is
\begin{eqnarray}
[p_i(a^0=1),p_i(a^1),\ldots,p_i(a^{n-1})]~,\label{eq:generator}
\end{eqnarray} where $p_i(x)$ is a polynomial with all elements in the $i$th row of $U$ as its coefficients, that is, $p_i(x)=\sum_{j=0}^{d-1}u_{ij}x^j$, and $a$ is a generator of $GF(2^m)$. In matrix form, we have $$U\cdot G=C,$$
where 
$$G=\left[\begin{array}{cccc}
1&1&\cdots&1\\
a^0&a^1&\cdots&a^{n-1}\\(a^0)^2&(a^1)^2&\cdots&(a^{n-1})^2\\
&&\vdots&\\
(a^0)^{d-1}&(a^1)^{d-1}&\cdots&(a^{n-1})^{d-1}\end{array}\right],$$ and $C$  is
the codeword vector with dimension $(\alpha\times n)$. Finally, the $i$th
column of $C$ is distributed to storage node $i$ for $1\le i\le n$. 

The  generator matrix $G$ of the RS code can be reformulated as
{\footnotesize \begin{eqnarray*}
\hspace{-0.5cm} G&=&\left[\begin{array}{cccc}
1&1&\cdots&1\\
a^0&a^1&\cdots&a^{n-1}\\(a^0)^2&(a^1)^2&\cdots&(a^{n-1})^2\\
&&\vdots&\\
(a^0)^{\alpha-1}&(a^1)^{\alpha-1}&\cdots&(a^{n-1})^{\alpha-1}\\
(a^0)^\alpha 1&(a^1)^\alpha 1&\cdots&(a^{n-1})^\alpha 1\\
(a^0)^\alpha a^0&(a^1)^\alpha a^1&\cdots&(a^{n-1})^\alpha a^{n-1}\\
(a^0)^\alpha(a^0)^2&(a^1)^\alpha(a^1)^2&\cdots&(a^{n-1})^\alpha(a^{n-1})^2\\
&&\vdots&\\
(a^0)^\alpha (a^0)^{\alpha-1}&(a^1)^\alpha(a^1)^{\alpha-1}&\cdots&(a^{n-1})^\alpha(a^{n-1})^{\alpha-1}
\end{array}\right]\\
&=&\left[\begin{array}{c}
\bar{G}\\
\bar{G}\Delta
\end{array}
\right]~,
\end{eqnarray*}}
where, $\bar{G}$ contains the first $\alpha$ rows in $G$ and $\Delta$ is a diagonal matrix with $(a^0)^\alpha,\ (a^1)^\alpha,\ (a^2)^\alpha,\ldots,\ (a^{n-1})^\alpha$ as diagonal elements. 
It is easy to see that the $\alpha$ symbols stored in storage node $i$ is
$$U\cdot \left[\begin{array}{c}
\gggg_i^T\\
(a^{i-1})^\alpha \gggg_i^T\end{array}\right]=A_1\gggg_i^T+(a^{i-1})^\alpha A_2\gggg_i^T,$$
where $\gggg_i^T$ is the $i$th column in $\bar{G}$.

A final remark is that each column in $G$ can be generated by knowing the index
of the column and the generator $a$. Therefore,  each storage node does not need to
store the entire $G$ to perform  exact-regeneration.

\subsection{Decoding for Data-Reconstruction}
\label{sec:decoding-msr-data}
The generator polynomial of the RS code encoded by~\eqref{eq:generator} has
$a^{n-d},a^{n-d-1},\ldots, a$ as roots~\cite{MOO05}. Without loss of
generality, we assume that the data collector retrieves encoded symbols from
$k$ storage nodes $j_0,\ j_1,\ldots,\  j_{k-1}$. First, the information
sequence $\m$ is recovered by the procedure given in~\cite{RAS11}.  Note that
the procedure in~\cite{ RAS11} requires that $(a^0)^\alpha,\ (a^1)^\alpha,\
(a^2)^\alpha,\ldots,\ (a^{n-1})^\alpha$ all be distinct. This can be guaranteed
if this code is over $GF(2^m)$ for $m\ge \lceil \log_2 n\alpha\rceil$. If the
recovered information sequence does not pass the CRC, then we need to
perform the error-erasure decoding. In addition to the received encoded symbols
from $k$ storage nodes, the data collector needs to retrieve the encoded symbols
from $d+2-k$ storage nodes of the remaining storage nodes. The data
collector then performs error-erasure decoding to obtain
$\tilde{C}$, the first $d$ columns of the codeword vector. Let ${\hat G}$ be
the first $d$ columns of $G$. Then the recovered  information sequence can be
obtained from
\begin{eqnarray}
\tilde{U}=\tilde{C}\cdot {\hat G}^{-1},\label{eq:U-MSR}
\end{eqnarray}
where $ {\hat G}^{-1}$ is the inverse of $ {\hat G}$ and it always exists. If
the recovered information sequence passes the CRC, it is done; otherwise,
two more symbols need to be retrieved.  The data collector continues the
decoding process until it successfully recover the correct information sequence
or no more storage nodes can be accessed.   In each step, the progressive
decoding that we proposed in~\cite{Han10-Infocom} is applied to reduce the
computation complexity.  Note that the RS code used is capable of correcting up
to $\lfloor (n-d)/2\rfloor$ errors.

The decoding  algorithm is summarized in
Algorithm~\ref{algo:reconstruction-MSR}. Note that, in practice,
Algorithm~\ref{algo:reconstruction-MSR} will be repeated $\beta$ times for each retrieved symbol when
$\beta > 1$. 
\begin{algorithm}[h]
\Begin {
The data collector randomly chooses $k$ storage nodes and retrieves encoded data,
$Y_{\alpha\times k}$;\\ 
Perform the procedure given in~\cite{RAS11} to recover $\tilde{\m}$;\\
\eIf{$CRCTest(\tilde{\m}) = SUCCESS$} {
\Return $\tilde{\m}$;}
{Retrieve $d-k$ more encoded data from remaining storage nodes and merge them into $Y_{\alpha\times d}$;\\
$i \leftarrow d$;\\
\While {$i \le n-2$} {
$i\leftarrow i+2$;\\
Retrieve two more encoded data from remaining storage nodes  and merge them into $Y_{\alpha\times i}$; \\
Perform progressive error-erasure decoding on each row in $Y$ to recover $\tilde{C}$;\\
Obtain $\tilde{U}$ by~\eqref{eq:U-MSR} and convert it to $\tilde{\m}$;\\

 \If{$CRCTest(\tilde{\m}) = SUCCESS$} {
\Return $\tilde{\m}$;
} 
}
\Return FAIL;
}
}
\caption{Decoding of MSR Codes for Data-Reconstruction}
\label{algo:reconstruction-MSR}
\end{algorithm}

\subsection{Verification for Regeneration}
\label{sec:verification-exact}
To verify whether the recovered data are the same as those stored in the failed
node, integrity check is needed. However, such check should be performed based
on information stored on nodes {\it other than} the failed node. We consider two
mechanisms for verification. 

In this first scheme, each storage node keeps the CRC checksums for the rest $n-1$
storage nodes. When the helper accesses $d$ surviving storage nodes, it also
asks for the CRC checksums for the failed node from them. Using the majority vote
on all receiving CRC checksums, the helper can obtain the correct CRC checksum if no
more than $\lfloor (d-1)/2\rfloor$ accessed storage nodes are compromised.  To
see the storage complexity of this scheme, let us take a numerical example.
Consider a [100, 20, 38] MSR code with $\alpha = 19, B = 4.18MB,\beta = 1000$.
The total bits stored in each node is then $19\times 11\times 1000=209000$
bits. If a 32-bit CRC checksum is added to each storage node, the redundancy is
$r(n-1)/\beta\alpha m = 32\times 99/209000 \approx 1.5\%$ and the extra bandwidth for
transmitting the CRC checksums is around $rd/\beta\alpha m = 1216/418000 \approx
0.3\%$.  Hence, both redundancy for storage and bandwidth are manageable for large $\beta$'s. 

When $\beta$ is  small, we adopt an error-correcting code to encode the $r$-bit
CRC checksum. This can improve the storage and bandwidth efficiency.  First we select the operating finite field $GF(2^{m'})$ such
that $2^{m'}\ge n-1$. Then an $[n-1,k']$ RS code with $k'=\lceil r/m'\rceil$ is
used to encode the CRC checksum. Note that this code is different from the RS
code used for  MSR data regenerating. In encoding the CRC checksum of a storage
node into $n-1$ symbols and distributing them to the $n-1$ other storage nodes,
extra $(n-1)m'$ bits are needed on each storage node. When the helper
accesses $d$ storage nodes to repair the failed node $i$, these nodes also send
out the symbols associated with the CRC checksum for node $i$. The helper  then
can perform error-erasure decoding to recover the CRC checksum. The maximum
number of compromised storage nodes among the accessed $d$ nodes that can be
handled by this approach is $\lfloor (d-k')/2\rfloor$ and the  extra bandwidth
is $dm'$.  Since $m'$  is much smaller than $n-1$ and $r$, the redundancy for
storage and bandwidth can be reduced.

\subsection{Decoding for Regeneration}
Let node $i$ be the failed node to be recovered.  During regeneration,  the
helper accesses $s$ surviving storage nodes, where $d\le s\le n-1$.
Without loss of generality, we assume that the storage nodes accessed are
$j_0$, $j_1$,$\ldots$, $j_{s-1}$.  Every accessed node takes the inner product
between its $\alpha$ symbols and 
\begin{eqnarray}
\gggg_i=[1,(a^{i-1})^1,(a^{i-1})^2,\ldots, (a^{i-1})^{\alpha-1}]~, \label{eq:g_i}
\end{eqnarray}
where $\gggg_i$ can be generated by index $i$ and the generator $a$, and sends
the resultant symbol to the helper. Since the MSR code is a linear code,
the resultant symbols transmitted, $y_{j_0},\ y_{j_1},\
y_{j_2},\ldots,\ y_{j_{s-1}}$, can be decoded to the codeword $\cc$, where
\begin{eqnarray*}
\cc&=&\gggg_i\cdot\left(U\cdot G\right)\\
&=&\left(\gggg_i\cdot U\right)\cdot G~,
\end{eqnarray*}
if $(n-s)+2e<n-d+1$, where $e$ is the number of errors among
the $s$ resultant symbols. Multiplying $\cc$ by the inverse of the first $d$
columns of $G$, i.e., $\hat{G}^{-1}$, one can recover $$\gggg_i\cdot U$$ which
is equivalent to $$\gggg_i\cdot [A_1\ A_2]=[\gggg_i\cdot A_1\ \gggg_i\cdot
A_2]~.$$
Recall that $\gggg_i$ is the transpose of $i$th column of $\bar{G}$,  the first $\alpha$ rows in $G$.  Since
$A_j$, for $j=1,2$, are symmetric matrices, $(\gggg_i A_j)^T=A_j\gggg_i^T$.  The
$\alpha$ symbols stored in the failed node $i$ can then be calculated as
\begin{eqnarray}
(\gggg_i A_1)^T+(a^{i-1})^\alpha(\gggg_i A_2)^T~.\label{eq:regeneration}
\end{eqnarray}

The progressive decoding procedure  in~\cite{Han10-Infocom} can be applied in
decoding $y_{j_0},\ y_{j_1},\ y_{j_2},\ldots,\ y_{j_{s-1}}$. First, the helper
accesses $d$ storage nodes and decodes $y_{j_0},\ y_{j_1},\ y_{j_2},\ldots,\
y_{j_{d-1}}$ to obtain $\cc$ and $\alpha$ symbols by \eqref{eq:regeneration}. Then, it
verifies the CRC checksum. If the CRC check is passed, the regeneration is
successful; otherwise, two more surviving storage nodes need to be accessed.
Then the helper decodes the received $y_{j_0},\ y_{j_1},\ y_{j_2},\ldots,\
y_{j_{d+1}}$ to obtain $\cc$ and recover $\alpha$ symbols.  The process repeats
until sufficient number of correctly stored data have been retrieved to recover
the failed node. Again, in practice, when $\beta > 1$, the decoding needs to be
performed $\beta$ times to recover  $\beta \alpha$ symbols before verifying the CRC checksum.  The data
regenerating algorithm is summarized in Algorithm~\ref{algo:regeneration-MSR}.

\begin{algorithm}[h]
\Begin {
Assume node $i$ is failed.

The helper randomly chooses $d$ storage nodes;\\
 Each chosen storage node combines its symbols as a $(\beta\times \alpha)$ matrix and multiply it by $\gggg_i$ in~\eqref{eq:g_i};\\
The helper collects these resultant vectors as a $(\beta\times d)$ matrix $Y$.\\
The helper obtains the CRC checksum for node $i$;\\

$i \leftarrow d$;\\
\Repeat{$i \ge n-2$} {
Perform progressive error-erasure decoding  on each row in $Y$ to recover $\tilde{C}$ (error-erasure decoding performs $\beta$ times);\\
$M =\tilde{C}\hat{G}^{-1}$, where $\hat{G}^{-1}$ is the inverse of the first $d$ columns of $G$;\\
Obtain the $\beta\alpha$ information symbols, $\sss$, from $M$ by the method given in~\eqref{eq:regeneration};\\
 \eIf{$CRCTest(\sss) = SUCCESS$} {
\Return $\sss$;
} {
$i\leftarrow i+2$;\\
The helper accesses two more remaining storage nodes;\\
Each chosen storage node combines its symbols as a $(\beta\times \alpha)$ matrix and multiply it by $\gggg_i$ given in~\eqref{eq:g_i};\\
The helper  merges the resultant vectors into $Y_{\beta\times i}$; 
}
}
\Return FAIL;
}
\caption{Decoding of MSR Codes for Regeneration}
\label{algo:regeneration-MSR}
\end{algorithm}

\section{Encoding and Decoding of Error-Correcting Exact-Regenerating Codes for the MBR Points}
\label{sect:mbr}
In this section we demonstrate that  by selecting the same RS codes as that for
MSR codes and designing a proper decoding procedure, the MBR codes
in~\cite{RAS11} can be extended to handle Byzantine failures.  Since the
verification procedure for MBR codes is the same as that of MSR codes, it is
omitted.

\subsection{Encoding}
\label{sec:encoding-mbr}
Let  the information sequence $\m=[m_0,m_1,\ldots, m_{B-1}]$ be arranged into
an information vector $U$ with size $\alpha\times d$ such that
\begin{eqnarray*}
u_{ij}=\left\{\begin{array}{cc}
u_{ji}=m_{k_1}&\mbox{ for }  i\le j\le k\\
u_{ji}=m_{k_2}&\mbox{ for } k+1\le i\le d,\ 1\le j \le k\\
0&\mbox{ otherwise}
\end{array}\right.~,
\end{eqnarray*}
where $k_1=(i-1)(k+1)-i(i+1)/2+j$ and $k_2=(i-k-1)k+k(k+1)/2+j$.
In matrix form, we have 
\begin{eqnarray}
\label{U-mbr}
U=\left[\begin{array}{cc}
A_1&A_2^T\\
A_2&\0\end{array}
\right]~,
\end{eqnarray}
where $A_1$ is a $k\times k$ matrix, $A_2$ a $(d-k)\times  k$ matrix, $\0$ is
the $(d-k)\times (d-k)$ zero matrix.  Both $A_1$ and $A_2$ are symmetric.  It
is clear that $U$ has a dimension $d\times d$ (or $\alpha\times d$).

We apply an $[n,d]$ RS code to encode each row of $U$. Let $p_i(x)$ be the
polynomial with all elements in $i$th row of $U$ as its coefficients. That is,
$p_i(x)=\sum_{j=0}^{d-1}u_{ij}x^j$. The corresponding codeword  of $p_i(x)$ is thus
\begin{eqnarray}
[p_i(a^0=1),p_i(a^1),\ldots,p_i(a^{n-1})]~.\label{eq:generator-2}
\end{eqnarray} 
Recall that  $a$ is a generator of $GF(2^m)$. In matrix form, we have
$$U\cdot G=C,$$
where 
\begin{eqnarray*}
G&=&\left[\begin{array}{cccc}
1&1&\cdots&1\\
a^0&a^1&\cdots&a^{n-1}\\
(a^0)^2&(a^1)^2&\cdots&(a^{n-1})^2\\
&&\vdots&\\
(a^0)^{k-1}&(a^1)^{k-1}&\cdots&(a^{n-1})^{k-1}\\
(a^0)^k&(a^1)^k&\cdots&(a^{n-1})^k\\
&&\vdots&\\
(a^0)^{d-1}&(a^1)^{d-1}&\cdots&(a^{n-1})^{d-1}\end{array}\right]~,
\end{eqnarray*}
 and $C$  is the codeword vector with dimension $(\alpha\times n)$. $G$ is called the generator matrix of the $[n,d]$ RS code. $G$ can be divided into two sub-matrices as
\begin{eqnarray*}
G=\left[\begin{array}{c}
G_k\\
B
\end{array}\right]~,
\end{eqnarray*}
where
\begin{eqnarray}
G_k
=\left[\begin{array}{cccc}
1&1&\cdots&1\\
a^0&a^1&\cdots&a^{n-1}\\
(a^0)^2&(a^1)^2&\cdots&(a^{n-1})^2\\
&&\vdots&\\
(a^0)^{k-1}&(a^1)^{k-1}&\cdots&(a^{n-1})^{k-1}\\
\end{array}\right]\label{G-k}
\end{eqnarray}
and 
$$B=\left[\begin{array}{cccc}(a^0)^{k}&(a^1)^{k}&\cdots&(a^{n-1})^{k}\\
&&\vdots&\\
(a^0)^{d-1}&(a^1)^{d-1}&\cdots&(a^{n-1})^{d-1}\end{array}\right]~.$$ Note that
$G_k$ is a generator matrix of the $[n,k]$ RS code and it will be used in the
decoding process for data-reconstruction.

\subsection{Decoding for Data-Reconstruction}
The generator polynomial of the RS code encoded by~\eqref{G-k}
has $a^{n-k},a^{n-k-1},\ldots, a$ as roots~\cite{MOO05}. Hence, the progressive
decoding scheme given in~\cite{Han10-Infocom} can be applied to decode the
proposed code if there are errors in the retrieved data. Unlike the decoding
procedure given in~\ref{sec:decoding-msr-data}, where an $[n,d]$ RS decoder is
applied, we need an $[n,k]$ RS decoder for MBR codes. 

Without loss of generality, we assume that the data collector retrieves encoded
symbols from $s$ storage nodes $j_0,\ j_1,\ldots,\  j_{s-1}$, $k\le s\le n$.
Recall that $\alpha = d$ in MBR. Hence, the data collector receives $d$ vectors
where each vector has $s$ symbols. Collecting the first $k$ vectors as $Y_{k}$
and the remaining $d-k$ vectors as $Y_{d-k}$. From \eqref{U-mbr},  we can view the
codewords in the last $d-k$ rows of $C$ as being encoded by $G_k$
instead of $G$.  Hence, the decoding procedure of $[n,k]$ RS codes can be applied on
$Y_{d-k}$ to recover the codewords in the last $d-k$ rows of $C$. Let
$\hat{G_k}$ be the first $k$ columns of $G_k$ and $\tilde{C}_{d-k}$ be the
recovered codewords in the last $d-k$ rows of $C$. $A_2$ in $U$ can  be
recovered as
\begin{eqnarray}
\tilde{A}_2=\tilde{C}_{d-k}\cdot\hat{G_k}^{-1}~.\label{A_2}
\end{eqnarray}
We then calculate $\tilde{A}_2^T\cdot B$ and only keep  the $j_0$th, $j_1$th,
$\ldots$, $j_{s-1}$th columns of the resultant matrix as $E$, and  subtract $E$
from $Y_k$:
\begin{eqnarray}
Y'_{k}=Y_k-E~.\label{Y_k}
\end{eqnarray}
Applying the RS decoding algorithm again on $Y'_{k}$ we can recover $A_1$ as
\begin{eqnarray}
\tilde{A}_1=\tilde{C}_{k}\cdot\hat{G_k}^{-1}~.\label{A_1}
\end{eqnarray}
 CRC checksum is computed on
the decoded information sequence to verify the recovered data. If CRC is
passed, the data reconstruction is successful; otherwise the progressive
decoding procedure is applied, where two more storage nodes need to be accessed
from the remaining storage nodes in each round until no further errors are
detected. The data-reconstruction algorithm is summarized in Algorithm~\ref{algo:reconstruction-MBR}.

\begin{algorithm}[h]
\Begin {
The data collector randomly chooses $k$ storage nodes and retrieves encoded data,
$Y_{d \times k}$;\\ 

$i \leftarrow d$;\\
\Repeat{$i \ge n-2$} {
Perform progressive error-erasure decoding on last $d-k$ rows in $Y$ to recover $\tilde{C}$ (error-erasure decoding performs $d-k$ times);\\
Calculate $\tilde{A}_2$ via~\eqref{A_2};\\
Calculate $\tilde{A}_2\cdot B$  and obtain $Y_k'$ via~\eqref{Y_k};\\
Perform progressive error-erasure decoding  on $Y_k'$ to recover the first $k$ rows in codeword vector (error-erasure decoding performs $k$ times);\\
Calculate $\tilde{A}_1$ via~\eqref{A_1};\\
Recover the information sequence $\sss$ from $\tilde{A}_1$ and $\tilde{A}_2$;\\
 \eIf{$CRCTest(\sss) = SUCCESS$} {
\Return $\sss$;
} {
$i\leftarrow i+2$;\\
Retrieve two more encoded data from remaining storage nodes  and merge them into $Y_{d\times i}$; \\}
}
\Return FAIL;
}
\caption{Decoding of MBR Codes for Data-Reconstruction}
\label{algo:reconstruction-MBR}
\end{algorithm}

\subsection{Decoding for Regeneration}
Decoding for regeneration with MBR is very similar to that with MSR. After
obtaining $g_i\cdot U$, we take its transposition. Since $U$ is symmetric, we
have $U^T=U$ and $$U^T\cdot g_i^T=U\cdot g_i^T~.$$ CRC check is performed on
all $\beta\alpha$ symbols. If the CRC check is passed, the $\beta\alpha$ symbols are the data stored in the failed node; otherwise, the progressive decoding procedure is applied.

\section{Analysis}
\label{sect:eval}

\begin{table*}[tbh]
\caption{Evaluation of MSR and MBR codes}
\label{tab:evaluation}
\begin{center}
\begin{tabular}{|c||c|c||c|c||}\hline
&\mc{2}{c||}{MSR code}&\mc{2}{c||}{MBR code}\\ \hline\hline
&Data-reconstruction&Regeneration&Data-reconstruction&Regeneration\\ \hline
Fault-tolerant capability against erasures& $n-k$& $n-d$& $n-k$&$n-d$\\ 
Fault-tolerant capacity against Byzantine faults&$\lfloor\frac{n-d}{2}\rfloor$&$\min\{\lfloor\frac{n-d}{2}\rfloor, \lfloor\frac{d-k'}{2}\rfloor\}$&$\lfloor\frac{n-k}{2}\rfloor$&$\min\{\lfloor\frac{n-d}{2}\rfloor, \lfloor\frac{d-k'}{2}\rfloor\}$\\ \hline
Security strength under forgery attack&$\min\{k,\lceil \frac{n-d+2}{2}\rceil\}-1$&$\min\{d,\lceil \frac{n-d+2}{2}\rceil\}-1$&$\min\{k,\lceil \frac{n-k+2}{2}\rceil\}-1$& $\min\{d,\lceil \frac{n-d+2}{2}\rceil\}-1$\\ \hline
Redundancy ratio on storage (bits)&$\frac{r}{mk\alpha-r}$&$\frac{(n-1)m'}{\beta\alpha m}$&$\frac{r}{m(kd-k(k-1)/2)-r}$& $\frac{(n-1)m'}{\beta\alpha m}$\\ \hline
Redundancy ratio on bandwidth (bits)&$\cdot$&$\frac{dm'}{\beta m d}=\frac{m'}{\beta m}$&$\cdot$& $\frac{dm'}{\beta m d}=\frac{m'}{\beta m}$\\ \hline
\end{tabular}
\end{center}
where $k'=\lfloor\frac{r}{m'}\rfloor$ and $m'=\lceil\log_2 (n-1)\rceil$
\end{table*}

In this section, we provide an analytical study of the fault-tolerant
capability, security strength, and storage and bandwidth efficiency of the
proposed schemes. 

\subsection{Fault-tolerant capability}
In analyzing the fault-tolerant capability, we consider two types of failures,
namely crash-stop failures and Byzantine failures. Nodes are assumed to fail
independently (as opposed in a coordinated fashion). In both cases, the
fault-tolerant capacity is measured by the maximum number of failures that the
system can handle to remain functional. 

\paragraph*{Crash-stop failure}
Crash-stop failures can be viewed as erasure in the codeword. Since at least
$k$ nodes need to be available for data-reconstruction, it is
easy to show that the maximum number of crash-stop failures that can be
tolerated in data-reconstruction is $n-k$. For regeneration,  $d$ nodes need to be accessed. Thus, the fault-tolerant
capability is  $n-d$. Note that since live nodes all
contain correct data, CRC checksum is also correct. 
\paragraph*{Byzantine failure}
In general, in RS codes, two additional correct code fragments are needed to
correct one erroneous code fragments. However, in the case of data
regeneration, the capability of the helper to obtain the correct CRC checksum also
matters. In the analysis, we assume that the error-correction code is used in
the process to obtain the correct CRC checksum. Data regeneration will fail if the
helper cannot obtain the correct CRC checksum even when the number of failed nodes
is less than the maximum number of faults the RS code can handle. Hence, we
must take the minimum of the capability of the RS code (in MBR and MSR) and the
capability to recover the correct CRC checksum. Thus, with MSR and MBR code,
$\lfloor \frac{n-d}{2}\rfloor$ and $\lfloor \frac{n-k}{2}\rfloor$ erroneous
nodes can be tolerated in data reconstruction. On the other hand, the
fault-tolerant capacity of MSR and MBR code for data regeneration are both
$\min\left\{\lfloor\frac{n-d}{2}\rfloor, \lfloor\frac{d-k'}{2}\rfloor\right\}$.

\subsection{Security Strength}
In analyzing the security strength, we consider forgery attacks, where
polluters~\cite{OGG11}, a type of Byzantine attackers, try to disrupt the
data-reconstruction and regenerating process by forging data cooperatively.
In other words, collusion among polluters are considered. We want to determine
the minimum number of polluters to forge the data in data-reconstruction and
regeneration. The security strength is therefore one less the number. Forgery
in data regeneration is useful when an attacker only has access to a small set
of nodes but through the data regeneration process ``pollutes" the data on
other storage nodes and thus ultimately leads to valid but erroneous data-reconstruction.

In data-reconstruction, for worst case analysis, we consider the security
strength such that only one row of $U$ is modified.\footnote{Due to symmetry in
$U$, most of the time, making changes on a row in $U$ results in changes on several
rows simultaneously.} Let the  polluters be $j_0,j_1,\ldots,j_{v-1}$, who can
collude to forge the information symbols.  Suppose that $\yy$ is the forged row
in $U$. Let ${\tilde \yy}=\yy+\uu$, where $\uu$ is the real  information
symbols in the   row of $U$. Then, according to the RS encoding procedure, we
have
\begin{eqnarray}
\yy G=({\tilde \yy}+\uu)G={\tilde \yy} G+\uu G=\vv+\cc,
\end{eqnarray}
where $\cc$ is the original data storage in storage nodes and $\vv$ is the
modified data must be made by the polluters. Let the number of nonzero symbols
in $\vv$ is $h$.  It is clear that $h\ge n-d+1$, where $n-d+1$ is the minimum
Hamming distance of the RS code, since $\vv$ must be a codeword. For worst-case
consideration, we assume that $h=n-d+1$. In order to successfully forge
information symbols, the attacker must compromise some storage nodes and make
them to store the corresponding encoded symbols in $\yy G$, the codeword
corresponding to the forged information symbols.  If the attacker compromises $
k$ storage nodes, then when the data collector happens to access these
compromised storage nodes, according to the decoding procedure, the attack can
forge the data successfully. Let the attacker compromise $b<k$ storage nodes.
According the decoding procedure, when $h-b=n-d+1-b\le \lfloor
\frac{n-d}{2}\rfloor$, where $\lfloor \frac{n-d}{2}\rfloor$ is the
error-correction capability of the RS code, the decoding algorithm still has
chance to decode the received vector to $\yy G$. Taking the smallest value of
$b$ we have $b=\lceil \frac{n-d+2}{2}\rceil$.  Hence, the security strength for
data-reconstruction is $\min\{k,\lceil \frac{n-d+2}{2}\rceil\}-1$ in MSR codes. Since the $[n,k]$ RS code is used in decoding for MBR codes, the security strength for
them becomes $\min\{k,\lceil \frac{n-k+2}{2}\rceil\}-1$.

Next we investigate the forgery attack on regeneration. Since computing the CRC checksum is a linear operation, there is no need for the attacker to break the CRC checksum for the failed node. It only needs to make the forged data with all zero  redundant bits. Hence, the security strength for regeneration is $\min\{d,\lceil \frac{n-d+2}{2}\rceil\}-1$.

It can be observe that CRC does not increase the security strength in forgery attack. By using hash value, the security strength can be increased since the operation to obtain hash value is non-linear. In this case, the attacker not only needs to obtain the original information data but also can forge hash value. Hence, the security strength can be increased to at least $k-1$  in data-reconstruction and at least $d-1$ for regeneration.\footnote{For  regeneration, the security strength is $\max\{d,  \min\{k',\lceil \frac{d-k'+2}{2}\rceil\}\}-1=d-1$ since $k'$ is usually less than $d$.}

\subsection{Redundancy Ratios on Storage and Bandwidth}
CRC checksums incur additional overhead in storage and bandwidth consumption.
The redundancy incurred for data-construction is $r$ bits, the size of CRC
checksum. Each information sequence is appended with the extra $r$ bits such that
it can be verified after reconstruction.  The  number of information bits is
$mk\alpha -r$ for MSR codes and $m(kd-k(k-1)/2)-r$ for MBR codes, respectively.
For regeneration, we assume that the $[n-1,k']$ RS code is used to distribute
the encoded CRC symbols to $n-1$ storage nodes, where
$k'=\lfloor\frac{r}{m'}\rfloor$ and $m'=\lceil\log_2 (n-1)\rceil$. Since each
storage node must store the encoded CRC symbols for other $n-1$ storage nodes,
the extra storage required for it is $(n-1)m'$ bits. The encoded data symbols
stored in each storage node is $\beta\alpha m$ bits.

The helper must obtain the correct CRC checksum for the failed node to verify the
correctness of the recovered data. The $d$ storage nodes accessed need to
provide their stored data associated with the CRC checksum of the failed node to
the helper.  Since each piece has  $m'$ bits, the total extra bandwidth is
$dm'$. The total bandwidth to repair the $\beta \alpha$ symbols stored in the
failed node is $\beta  md$. 

Table~\ref{tab:evaluation} summarizes the quantitative results of fault-tolerate
capability, security strength, and redundancy ratio of the MSR and MBR codes. 
\section{Related Work}
\label{sect:related}
Regenerating codes were introduced in the pioneer works by Dimakis {\it et al.}
in~\cite{DIM07,DIM10}. In these works, the so-called cut-set bound was derived
which is the fundamental  limit for designing regenerating codes. In these
works, the data-reconstruction and regeneration problems were formulated as a
multicast network coding problem. From the cut-set bounds between the source
and the destination, the parameters of the regenerating codes were shown to
satisfy \eqref{main-inequality}, which reveals the  tradeoff between storage
and repair bandwidth. Those parameters satisfying the cut-set bound with
equality were also derived.

The regeneration codes with parameters satisfying the cut-set bound with equality
were proposed in~\cite{WU07,WU10}. In~\cite{WU07}  a deterministic construction
of the generating codes with $d=n-1$ was presented. In~\cite{WU10}, the network
coding approach was adopted to design the generating codes. Both constructions
achieved functional regeneration but exact regeneration.

Exact regeneration was considered in~\cite{CUL09,WU09,RAS09}. In~\cite{CUL09},
a search algorithm was proposed to search for exact-regenerating MSR codes with
$d=n-1$; however, no systematic construction method was provided.
In~\cite{WU09}, the MSR codes with $k=2,d=n-1$ were constructed by using the
concept of interference alignment, which was borrowed from the context of
wireless communications. A drawback of this approach is that it 
operates on a finite field with a large size. In~\cite{RAS09}, the authors
provided an explicit method to construct the MBR codes with $d=n-1$. No
computation is required for these codes during the regeneration of a failed
node. Explicit construction of the MSR codes with $d=k+1$ was also provided;
however, these codes can perform exact regeneration only for a subset of failed storage
nodes.

In~\cite{RAS10}, the authors proved that exact regeneration is impossible for
MSR codes with $[n,k, d<2k-3]$ when $\beta=1$. Based on interference alignment
approach, a code construction was provided for MSR codes with $[n=d+1,k,d\ge
2k-1]$.  In~\cite{RAS11}, the explicit constructions for optimal MSR codes with
$[n,k,d\ge 2k-2]$ and optimal MBR codes were proposed. The construction was
based on the product of tow matrices: information matrix and encoding matrix.
The information matrix (or its submatrices) is symmetric in order to have
exact-regeneration property.

The problem of security on regenerating codes were considered
in~\cite{PAW11,OGG11}. In~\cite{PAW11}, the authors considered the security
problem against eavesdropping and adversarial attackers during the regeneration
process. They derived upper bounds on the maximum amount of information that
can be stored safely. An explicit code construction was given for $d=n-1$ in
the bandwidth-limited regime. The problem of Byzantine fault tolerance for
regenerating codes was considered in~\cite{OGG11}. The authors studied the
resilience of regenerating codes which support multi-repairs. By using
collaboration among newcomers (helpers), upper bounds on the resilience
capacity of regenerating codes were derived. Even though our work also deals with the Byzantine failures, it does not need to have multiple helpers to recover the failures.

The progressive decoding technology for distributed storage was first
introduced in~\cite{Han10-Infocom}. The scheme retrieved just enough data from
surviving storage nodes to recover the original data in the presence of
crash-stop and Byzantine failures. The decoding was performs  incrementally
such that both communication and computation cost are minimized.

\section{Conclusions}
\label{sect:conclusion}
In this paper, we considered the problem of exact  regeneration with error
correction capability for Byzantine fault tolerance in distributed storage
networks. We showed the Reed-Solomon codes combined with CRC checksum can be
used for both data-reconstruction and regenerating, realizing MSR and MBR in
the later case. Progressive decoding can be applied in both applications to reduce
the computation complexity in presence of erroneous data. Analysis on the fault
tolerance, security, storage and bandwidth overhead shows that the proposed
schemes are effective without incurring too much overhead. 



\end{document}